\documentclass[conference]{IEEEtran}
\IEEEoverridecommandlockouts
\usepackage{todonotes}

\usepackage{cite}
\usepackage{amsmath,amssymb,amsfonts}
\usepackage{algorithmic}
\usepackage{graphicx}
\usepackage{textcomp}
\usepackage{xcolor}
\usepackage{url}
\usepackage{hyperref} 
\usepackage{tikz}
\usetikzlibrary{patterns, positioning}
\usepackage{float}

\usepackage{booktabs}
\usepackage[table]{xcolor}

\def\BibTeX{{\rm B\kern-.05em{\sc i\kern-.025em b}\kern-.08em
    T\kern-.1667em\lower.7ex\hbox{E}\kern-.125emX}}
\begin{document}

\newcommand{\CPP}{C\nolinebreak\hspace{-.05em}\raisebox{.4ex}{\tiny\bf +}\nolinebreak\hspace{-.10em}\raisebox{.4ex}{\tiny\bf +}}
%\title{Scaling MPI and Asynchronous Many-Task Runtimes with the FleCSI Framework}
\title{Radiation Hydrodynamics at Scale: Comparing MPI and Asynchronous Many-Task Runtimes with FleCSI}
\author{\IEEEauthorblockN{1\textsuperscript{st} Alexander Strack}
\IEEEauthorblockA{\textit{Institute for Parallel} \\ \textit{and Distributed Systems} \\
\textit{University of Stuttgart}\\
Stuttgart, Germany \\
alexander.strack@ipvs.uni-stuttgart.de}
\and
\IEEEauthorblockN{2\textsuperscript{nd} Hartmut Kaiser}
\IEEEauthorblockA{\textit{Center of Computation and Technology} \\
\textit{Louisiana State University}\\
Baton Rouge, LA, USA \\
hkaiser@cct.lsu.edu}
\and

\IEEEauthorblockN{3\textsuperscript{th} Dirk Pfl\"uger}
\IEEEauthorblockA{\textit{Institute for Parallel} \\ \textit{and Distributed Systems} \\
\textit{University of Stuttgart}\\
Stuttgart, Germany \\
dirk.pflueger@ipvs.uni-stuttgart.de}
}

\maketitle
\begin{abstract}
Writing efficient distributed code remains a labor-intensive and complex endeavor.
To simplify application development, the Flexible Computational Science Infrastructure (FleCSI) framework offers a user-oriented, high-level programming interface that is built upon a task-based runtime model.
Internally, FleCSI integrates state-of-the-art parallelization backends, including MPI and the asynchronous many-task runtimes (AMTRs) Legion and HPX, enabling applications to fully leverage asynchronous parallelism.

In this work, we benchmark two applications using FleCSI’s three backends on up to 1024 nodes, intending to quantify the advantages and overheads introduced by the AMTR backends.
As representative applications, we select a simple Poisson solver and the multidimensional radiation hydrodynamics code HARD.
In the communication-focused Poisson solver benchmark, FleCSI achieves over 97\% parallel efficiency using the MPI backend under weak scaling on up to 131072 cores, indicating that only minimal overhead is introduced by its abstraction layer.
While the Legion backend exhibits notable overheads and scaling limitations, the HPX backend introduces only marginal overhead compared to MPI+Kokkos.
However, the scalability of the HPX backend is currently limited due to the usage of non-optimized HPX collective operations.

In the computation-focused radiation hydrodynamics benchmarks, the performance gap between the MPI and HPX backends fades.
On fewer than 64 nodes, the HPX backend outperforms MPI+Kokkos, achieving an average speedup of 1.31 under weak scaling and up to 1.27 under strong scaling.
For the hydrodynamics-only HARD benchmark, the HPX backend demonstrates superior performance on fewer than 32 nodes, achieving speedups of up to 1.20 relative to MPI and up to 1.64 relative to MPI+Kokkos.

These results underscore two key insights: FleCSI maintains excellent scalability and efficiency despite its abstraction layer, and the AMTR backends, while introducing additional overheads, exhibit strong potential to enhance performance in complex and asynchronous workloads.
\end{abstract}


\section{Introduction}\label{sec:introduction}

Scientific computing is evolving in a direction that enables us to capture a greater portion of real-world phenomena in simulations through more sophisticated and complex models. 
Such simulations are typically executed on large supercomputers. 
Given the recent trend towards many-core computing nodes, which can also include accelerators, programmers are required to be fluent in advanced parallelization tools, such as MPI~\cite{Mpi2021}, CUDA~\cite{nvidia2025_cuda}, and Kokkos~\cite{Trott2022_kokkos}, to leverage modern computing architectures.
The Flexible Computational Science Infrastructure (FleCSI)~\cite{Bergen2022_flecsi} provides a clean, high-level interface using modern C\texttt{++} syntax.
Additionally, the code is highly portable and can target all modern hardware via a Kokkos~\cite{Trott2022_kokkos} backend. 
Distributed communication is completely abstracted, and a high-level task-based programming model allows for parallelization with asynchronous many-task runtimes (AMTRs), such as HPX~\cite{Kaiser2020_hpx} and Legion~\cite{Bauer2012_legion}.
FleCSI supports synchronous execution with MPI as well.
The abstractions introduced by FleCSI itself and by the AMTR backends add overhead that can impact performance.
In benchmarks involving AMTRs, workloads are typically selected where AMTRs are expected to improve performance.
As a result, the actual abstraction overhead can not be accurately quantified. 
Thus, in this work, we specifically pick one application where the classical parallelization approach based on MPI+X is better suited.
This allows us to make the bare cost of using AMTRs over optimized MPI within FleCSI visible.
Furthermore, we pick a more complex application better suited for asynchronous parallelization.
Based on this approach, we formulate the following research questions:
\begin{itemize}
\item How do the general abstractions introduced by FleCSI affect the scalability of basic MPI-based parallelization?
\item What is the respective overhead introduced by the two AMTR backends compared to the synchronous MPI and MPI+Kokkos backends?
\item Can the AMTR task management overhead be effectively hidden in more complex applications, and can AMTRs even improve performance?
\end{itemize}

Therefore, we focus on evaluating the scalability and performance of the three FleCSI backends on CPU-only nodes.
GPU acceleration is not considered in this work.
Nevertheless, the FleCSI framework supports targeting accelerators via Kokkos in conjunction with all backends.
Heterogeneous computing, however, introduces additional communication overhead between the CPU and GPU, which we exclude to isolate the overheads induced by the AMTRs.

We chose two scientific computing applications from the field of computational fluid dynamics (CFD) as benchmarks.
Our first application is a red-black Gauss-Seidel pressure solver for the Poisson equation that offers a communication-focused benchmark problem.
While not containing complex physics or higher-accuracy numerical methods, it exposes any overheads introduced by the AMTRs.
The second application is the radiation hydrodynamics code HARD~\cite{Loiseau2025_hard}, which resembles a computationally intensive real-world workload.
HARD supports the simulation of one, two, and three-dimensional problems.
To evaluate FleCSI with these two applications, we perform strong and weak scaling tests on up to 1024 nodes on the Chicoma supercomputer.

Our main contributions in this work include:
\begin{itemize}
\item The first evaluation of the FleCSI framework at a large scale across all available FleCSI backends with multiple applications,
\item A comparative analysis of the MPI, Legion, and HPX backends, all executing the same C\texttt{++} code implemented using FleCSI, and
\item The development of a new benchmark mode for the FleCSI Poisson example, enabling asynchronous timing and streamlined benchmarking of the various FleCSI backends.
\end{itemize}

The remainder of this work is as follows:
In Section \ref{sec:related_work}, we review related work regarding FleCSI and AMTRs.
Then, in Section \ref{sec:framework}, we introduce the FleCSI framework and provide an overview of the different parallelization backends available in FleCSI. 
In Section \ref{sec:application}, we introduce the applications we use in this work for benchmarking FleCSI.
The results of our benchmarks on the Chicoma supercomputer are discussed in Section \ref{sec:results}.
Lastly, in Section \ref{sec:conclusion}, we conclude our work and give an outlook on future work.

\section{Related work}\label{sec:related_work}

Development of FleCSI started as early as 2016~\cite{Bergen2016_flecsi_1}. 
FleCSI~2.0 was introduced in~\cite{Bergen2022_flecsi} and includes significant changes from the original implementation. 
The FleCSI framework provides a unified interface for both shared-memory and distributed memory programming. 
The Uintah framework~\cite{Holmen2021_uintah} adopts a similar approach, providing a unified interface and a task-based runtime. However, while task dependencies are implicitly defined within FleCSI, users must explicitly allow concurrent execution in Uintah. Furthermore, Uintah employs a static resource mapping while FleCSI supports dynamic resource mapping of computational tasks.
Another similar approach takes the Multi-Processor Computing runtime (MPC)~\cite{Perache2009_mpc-mpi}. 
While also supporting an MPI+X programming model, resource mapping is static to MPI ranks.

FleCSI enables users to utilize various backends for distributed computing, including a synchronous MPI backend, as well as two asynchronous backends that leverage Legion~\cite{Bauer2012_legion} and HPX~\cite{Kaiser2020_hpx}. 
We refer to Section \ref{sec:framework} for a more detailed introduction to the different backends. 
A Kokkos backend ensures performance portability~\cite{Trott2022_kokkos} and integrates seamlessly with FleCSI’s distributed backends.
In addition, FleCSI has built-in support for the Caliper toolbox~\cite{Boehme2016_caliper}. 
Caliper enables easy performance measurement and profiling of applications built on top of FleCSI.

Applications of FleCSI include FleCSPH~\cite{Loiseau2020_flecsph}, Moya~\cite{Herring2025_flecsi_hpx}, and HARD~\cite{Loiseau2025_hard}. FleCSPH implements a smoothed-particle hydrodynamics solver leveraging tree-based data structures. In contrast, HARD relies on a regular grid but implements radiation hydrodynamics with a flux-limited diffusion approach~\cite{Levermore1981_fld_original, Turner2001_fld_numerical}. 

A variety of asynchronous tasking runtimes exist, and although they share certain similarities, each exhibits unique features.
We refer to Thoman et al.~\cite{Thoman2018_taxonomy} for a general taxonomy and to Schuchart et al.~\cite{Schuchart2025} for an AMTR comparison survey. 
With respect to performance, the Task Bench framework enables the comparison of different AMTRs for several algorithms~\cite{Slaughter2020_taskbench, Wu2023_taskbench_hpx}. 
However, in Task Bench, each AMTR implements the algorithms differently. 
With FleCSI, we can run the same code with different backends.

\section{Software framework}\label{sec:framework}

FleCSI provides a set of programming tools~\cite{Bergen2022_flecsi}, lowering the entry barrier for parallel programming.
Specifically, the FleCSI core framework provides a set of distributed data topologies: 
The \texttt{topo::narray} topology implements an arbitrary n-dimensional array that is well-suited for structured grid applications, such as Eulerian hydrodynamics; 
The \texttt{topo::unstructured} topology is a graph data structure that is designed for unstructured grid applications such as Finite Element or Finite Volume Methods; and, for applications that utilize tree-based algorithms, such as smoothed-particle hydrodynamics, FleCSI offers the \texttt{topo::ntree} topology.
The FleCSI runtime manages the communication between dependent parts of these distributed topologies. 
This allows developers to directly access data from a ghost or boundary layer without explicit communication. 
In addition to these low-level data structures, FleCSI specializations enable the implementation of a set of higher-level data structures, such as multi-dimensional meshes.

Control flow in FleCSI is defined through a control model that contains a hierarchy of control points, actions, tasks, and kernels. This structure allows the user to write portable, extensible application code that will align with different system architectures and can identify and exploit concurrency.
For a more detailed description of the FleCSI control, data, and execution models, we refer to~\cite{Bergen2022_flecsi}.

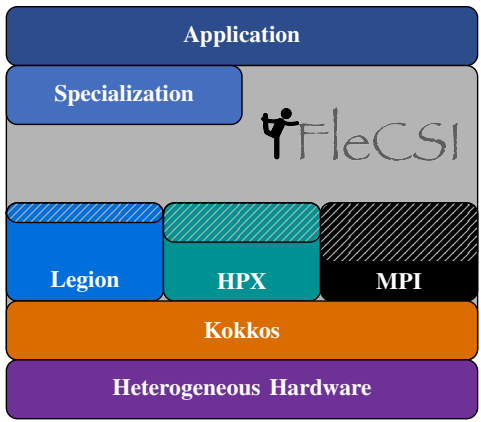
\begin{figure}
    \centering
    
\resizebox{0.35\textwidth}{!}{%
        \begin{tikzpicture}[
                font=\LARGE\bfseries,
                line width=1pt,
                rounded corners=8pt
                ]
        
        % ---------- Dimensions ----------
        \def\W{12}     % total width
        \def\H{12}     % total height
        
        % ---------- Colors ----------
        \definecolor{appblue}{RGB}{45,76,140}
        \definecolor{specblue}{RGB}{70,110,190}
        \definecolor{flecsibg}{RGB}{180,180,180}
        
        \definecolor{legion}{RGB}{0,110,219}
        \definecolor{hpx}{RGB}{0,146,146}
        \definecolor{mpi}{RGB}{0,0,0}
        
        \definecolor{kokkos}{RGB}{219,109,0}
        \definecolor{hardware}{RGB}{110,50,150}
        
        % ---------- Top: Application ----------
        \draw[fill=appblue] (0,10.5) rectangle (\W,12);
        \node[white] at (\W/2,11.25) {Application};

        % ---------- FleCSI area ----------
        \draw[fill=flecsibg] (0,3) rectangle (\W,10.5);
        \node at (9,8.75) {\includegraphics[width=5cm]{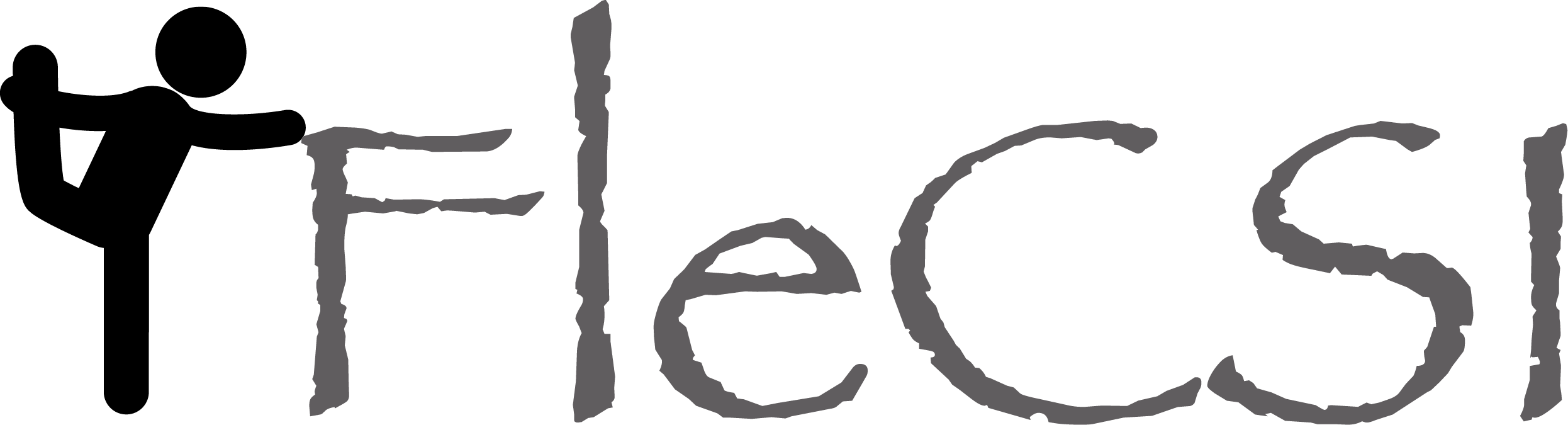}};
        
        % ---------- Specialization ----------
        \draw[fill=specblue] (0,9) rectangle (6,10.5);
        \node[white] at (3,9.75) {Specialization};

        % ---------- Runtime layer (Legion / HPX / MPI) ----------
        % Solid boxes
        \draw[fill=legion] (0,4.5) rectangle (4,7);
        \draw[fill=hpx]    (4,4.5) rectangle (8,7);
        \draw[fill=mpi]    (8,4.5) rectangle (12,7);
        
        \node[white] at (2,5) {Legion};
        \node[white] at (6.0,5) {HPX};
        \node[white] at (10,5) {MPI};
        
        % Pattern overlays
        \draw[pattern=north east lines, pattern color=flecsibg]
          (0,6.5) rectangle (4,7);
        
        \draw[pattern=north east lines, pattern color=flecsibg]
          (4,6) rectangle (8,7);
        
        \draw[pattern=north east lines, pattern color=flecsibg]
          (8,5.5) rectangle (12,7);
        
        % ---------- Kokkos ----------
        \draw[fill=kokkos] (0,3) rectangle (\W,4.5);
        \node[white] at (\W/2,3.75) {Kokkos};
        
        % ---------- Hardware ----------
        \draw[fill=hardware] (0,1.5) rectangle (\W,3);
        \node[white] at (\W/2,2.25)
          {Heterogeneous Hardware};
        
        % ---------- Borders ----------
        %\draw[line width=1pt] (0,1.5) rectangle (\W,12);
    \end{tikzpicture}
    }
    \caption{FleCSI backend stack}
    \label{fig:backends}
\end{figure}

In this work, we focus on the different FleCSI backends. 
The FleCSI backend stack is illustrated in Figure \ref{fig:backends}.
The tasks defined in the FleCSI control model can be executed with three different backends. 
The choice of the backend is determined at compile time. 
If built against the MPI backend, FleCSI runs through the task graph sequentially. 
In contrast, the Legion and HPX backends can analyze dependencies between tasks and run multiple independent tasks concurrently.  
In FleCSI, dependencies between tasks are implicitly defined by data access rights (\textit{privileges} in FleCSI's nomenclature) on the fields that are passed into a task.
This closely resembles the Legion approach to dependency management (see Subsection \ref{sec:legion}). 
The HPX backend automatically maps these implicit dependencies onto a future-based dependency representation~\cite{Herring2025_flecsi_hpx}.

For computation kernels, FleCSI provides a thin \texttt{for\_all} abstraction on top of \texttt{Kokkos::parallel\_for} that preserves the FleCSI range model and enforces memory consistency for collective operations.
As a result, FleCSI gains access to all of the portability advantages of Kokkos while providing an intersection between distributed and shared-memory parallelization. 
On CPUs, FleCSI's \texttt{for\_all} can use OpenMP or HPX through Kokkos.
On GPUs, it can use CUDA, ROCm, or SYCL through Kokkos. 
Thus, FleCSI can target the hardware of all major vendors and is well-equipped for modern heterogeneous computing environments.

Since the distributed backends are the primary focus of this work, we provide a brief introduction to MPI, followed by Legion and HPX in the subsequent subsections.

\subsection{MPI}

The Message Passing Interface (MPI)~\cite{Mpi2021} was introduced in 1994. 
It has become the de facto standard for communication among distributed memory entities. 
As its name implies, MPI enables data exchange through the explicit sending and receiving of messages. 
It supports both blocking and non-blocking point-to-point communication between MPI ranks.
However, for complex parallel algorithms, relying solely on point-to-point communication can become unwieldy, resulting in cluttered code and an increased risk of errors.
To address these issues, MPI offers a range of collective operations that match predefined communication patterns. 
These collectives operate in groups of ranks organized within a communicator. 
It is important to note that MPI is a specification that defines syntax and functionality, rather than a concrete implementation.
Popular implementations that are broadly available include OpenMPI~\cite{Gabriel2004_openmpi} and MPICH~\cite{Groop1999_mpich}. 
Some hardware vendors provide their own or specifically tailored MPI flavors, such as Cray MPICH~\cite{cray-mpich} by HPE Cray.

The MPI backend in FleCSI was initially developed as a risk mitigation strategy to provide a serial implementation of the FleCSI programming model.
Nevertheless, FleCSI's data privileges allow for a very clean MPI implementation that maintains memory coherence without explicit communication semantics, i.e., dependencies are inferred from field access rights.
FleCSI's collectives interface uses the respective MPI collectives directly, with point-to-point communication for ghost and boundary layer exchanges.
Because tasks are executed sequentially in the MPI backend, there is very little overhead incurred for runtime dependency analysis, task creation, and scheduling.
In contrast, HPX and Legion need dynamic analysis for out-of-order task execution.

\subsection{Legion}\label{sec:legion}

Legion is a parallel programming system developed at Stanford University~\cite{Bauer2012_legion}. 
Legion introduces an abstraction that separates computation from execution details.
Data is organized into logical regions.
A local region is a user-defined collection that mirrors complex data structures. 
Programmers express tasks in terms of these regions, declaring how each task accesses data. 
Legion provides a default \textit{mapper} to determine where and how tasks and data are placed on the available resources.
It is also possible to write custom mappers for specific applications or system architectures.
 
Legion builds on Realm~\cite{Treichler2014_realm} as a low-level resource manager, controlling, among other things, memory allocations and communication. 
Realm currently supports a variety of network transports, including MPI~\cite{Mpi2021}, GASNet~\cite{Bonachea2017_gasnet}, and UCX~\cite{Shamis2015_ucx}. 
Realm can be controlled with command-line arguments for various runtime parameters.
The Regent~\cite{Slaugther2015_regent} domain-specific language builds on top of Legion. 
It simplifies the Legion model to write high-level code and express the same behavior in fewer lines of code.
An example application of Regent is the unstructured mesh code PENNANT~\cite{Ferenbaugh2014_pennant}. % cut last 3 sentences if necessary
Software development using the Legion runtime was compared with MPI in~\cite{Mirchandaney2024_spinifel_legion}, utilizing the single-particle imaging code SpiniFEL~\cite{Chang2021_spinifel}.

FleCSI implements a partially customized mapper for the Legion backend, with support for additional customization in the specialization and application layers.
Because the Legion backend supports dynamic mapping of data and tasks onto system compute resources, additional options are possible for the runtime configuration of Realm.
For example, Legion supports multiple data partitions (\textit{colors}) for each process.
While this feature is logically more complicated, it has the advantage that fewer processes need to be managed by the runtime to support the same degree of distributed-memory parallelism.
Such a configuration eliminates redundant work in the Legion backend, as only one runtime process per address space is required for dependency analysis and scheduling.
Another performance optimization of Legion is the availability of a \textit{tracing} interface that reduces analysis overhead by identifying interactive sections of code where the same runtime analysis can be used for each iteration.
This can lead to significant performance improvements, although it is often difficult to overcome Legion's runtime overhead for applications that are primarily bulk synchronous.

\subsection{HPX}\label{sec:hpx}

The HPX~\cite{Kaiser2020_hpx} runtime was first released in 2008.
HPX is a C\texttt{++} standard-conforming library focused on enabling high-performance concurrency and parallelism. 
Its API adheres to the ISO C\texttt{++}11/14/17/20 standards and aligns with the programming guidelines of the Boost C\texttt{++} libraries. 
One of HPX's distinguishing features is its fully asynchronous execution model, achieved through lightweight HPX futures.
This approach eliminates the implicit synchronization barriers often found in traditional message-passing or fork-join approaches. 
Leveraging an active global address space (AGAS), HPX extends its capabilities to distributed computing~\cite{Heller2019_agas}. 
Communication is abstracted with active messages called \textit{parcels}.
HPX can use different parcelports to transmit these active messages. 
Options include TCP~\cite{Cerf1974_tcp}, MPI~\cite{Mpi2021}, and the Lightweight Communication Interface (LCI)~\cite{Yan23_lci}.
The LCI parcelport shows significant performance advantages over MPI and TCP~\cite{Strack2025_parcelports}.
HPX is designed for portability across a wide range of processor architectures, including x86, ARM~\cite{Diehl2023_arm}, and RISC-V~\cite{Diehl2023_riscv, Strack2026_riscv}.
HPX also has direct support for CUDA, ROCM, Kokkos, and SYCL to target accelerator architectures~\cite{Daiss2022_hpx_kokkos, Daiss2023_hpx_sycl}.  
The flagship application of HPX is the astrophysics code Octo-Tiger~\cite{Marcello2021_octotiger}.
Other applications of HPX include the Gaussian process regression library GPRat~\cite{Helmann2025_gprat} and the fast Fourier transform tool HPX-FFT~\cite{Strack2024_hpxfft}.

The HPX backend identifies execution dependencies among FleCSI tasks by examining specific annotations that users provide via the task's arguments.
It then automatically builds a future-based dependency tree to expose these execution dependencies to the scheduler.
Each FleCSI task is scheduled as an HPX task to ensure it executes only after all relevant FleCSI data it depends on has been updated, in accordance with the access right annotations specified for that task's corresponding argument.
This ensures that all tasks run as early as possible and with as much concurrency as possible. In particular, independent FleCSI tasks may run concurrently if sufficient compute resources are available. 

A feature not yet available during benchmarking is the communicator recycling added to the HPX backend in~\cite{Herring2025_flecsi_hpx}, which improves scaling for more complex applications.
Furthermore, we encountered some minor compatibility issues between HPX and Kokkos within FleCSI. 
To utilize the \texttt{for\_all} abstraction, we replaced the \texttt{Kokkos::parallel\_for} with an \texttt{hpx::for\_each} parallel loop as a temporary workaround for CPUs.

\section{Application}\label{sec:application}

In contrast to generic benchmarks, evaluating actual scientific applications that utilize FleCSI provides a more representative indication of real-world performance.
In scientific computing, CFD serves as a head term for a variety of methods that aim to solve physical equations modeling fluids or gases. 
For this work, we chose two CFD applications of different complexity that are introduced in the following subsections.

\subsubsection{Poisson solver}

We start with a standard problem commonly taught in CFD and numerical methods courses, namely, solving the Poisson equation. 
The Poisson equation, given by
\begin{equation}{\label{eq:poisson}}
    \Delta p = f,
\end{equation}
is an elliptic partial differential equation where $f$ is typically given, $\Delta$ denotes the Laplace operator and $p$ is the unknown. 
In two dimensions, Equation (\ref{eq:poisson}) can be written as 
\begin{equation}{\label{eq:twod}}
    \Bigg( \frac{\partial}{\partial x^2} + \frac{\partial}{\partial y^2}\Bigg)p(x,y) = f(x,y).
\end{equation}
For our application, we settle for a simple finite difference discretization on a regular, structured mesh.
Thus, the Laplacian of $p$ can be approximated as
\begin{align}{\label{eq:discrete}}
    \Delta p &= \Bigg( \frac{\partial}{\partial x^2} + \frac{\partial}{\partial y^2}\Bigg)p \\
             &\approx \frac{p_{i-1,j} - 2p_{i,j} + p_{i+1,j}}{\partial x^2} + \frac{p_{i,j-1} - 2p_{i,j} + p_{i,j+1}}{\partial y^2}.
\end{align}
Inserting Equation (\ref{eq:discrete}) into Equation (\ref{eq:twod}) results in the system of linear equations 
\begin{equation}{\label{eq:sle}}
    A\cdot p=f\,,
\end{equation}
where $A$ is a sparse matrix. 
Thus, an iterative solver can accelerate solving Equation (\ref{eq:sle}).
In this example, we choose the Gauss-Seidel method (GSM)~\cite{Saad2003_sparse_methods}. 
As iterations in the GSM depend on previously computed values, the method cannot be directly parallelized.
However, the mesh can be divided into a checkerboard pattern. 
In literature, this approach is referred to as red-black GSM~\cite{Saad2003_sparse_methods}.
The red mesh only depends on values from the black mesh, and vice versa. 
Thus, computations in the respective sub-meshes can be parallelized.

Using the task-based implementation with the FleCSI framework, we obtain a solve task that performs one time step and is repeated until convergence.
Within this task, 50 iterations of the red-black GSM are performed.
Thus, the resulting task graph is a pipeline of dependent solve tasks.
Through a mesh data structure based on the distributed \texttt{topo::narray} topology provided by FleCSI, the data is equally distributed across all processes.
As a result, the workload is homogeneous and computationally lightweight.  
The combination of these properties makes the Poisson application best suited for the synchronous MPI backend.
This allows us to highlight the overheads of the AMTR backends within FleCSI. 

A basic Poisson application is provided within FleCSI and may serve as a standalone example to facilitate further development.
In this work, we extend the example by introducing a benchmark mode that enables asynchronous timing.
The original implementation employed only high-level synchronous timing, which introduces implicit synchronization barriers.
Such barriers can distort the measured runtimes of the AMTR backends by adding artificial synchronization.
The new benchmark mode relocates timing instrumentation into lower-level FleCSI tasks, which are executed asynchronously, thereby ensuring barrier-free performance measurements.
In addition, the benchmark mode disables I/O operations and generates an output file containing the recorded timing data.

\subsubsection{HARD}

As a more complex application, representative of a real-world FleCSI use case, we include the HARD library~\cite{Loiseau2025_hard}. 
HARD enables the simulation of compressible hydrodynamics coupled with radiative diffusion.
It solves the radiation hydrodynamics equations with a flux-limited diffusion approximation~\cite{Levermore1981_fld_original}. 
HARD operates on a regular mesh and uses operator splitting to separate the advective and non-advective components of radiative transfer.
The advective part is solved using a finite-volume method, with fluxes computed via an approximate Riemann solver.
To achieve higher-order spatial accuracy, HARD employs fifth-order WENO5-Z reconstruction~\cite{Borges2008_hard_wenoz}.
For the hydrodynamics component, HLL fluxes are used~\cite{Harten1983_hard_hll_flux}, while the radiative component employs Lax-Friedrichs fluxes~\cite{Rusanov1962_hard_lax_flux}. 
Second-order temporal accuracy is achieved using Heun's method. 
The radiative diffusion is solved using a geometric multigrid solver with Jacobi smoothing~\cite{Briggs2000_multigrid}. 
For improved robustness and scalability, we only use the Jacobi solver in this work.

The combination of multiple higher-order numerical methods and the tight coupling of multidimensional hydrodynamics and radiation leads to a complex and computationally intensive task graph.
FleCSI’s task-based programming model allows for expressing these dependencies naturally while automatically managing all required communication.

As a benchmark scenario, we use a three-dimensional Rankine–Hugoniot problem~\cite{Rankine1870_rk, Hugoniot1887_rk}, which is particularly challenging to compute due to the presence of sharp discontinuities that require higher-order numerical methods.
The radiative component can be completely disabled, allowing HARD to act as a pure hydrodynamics solver. 
Additionally, HARD provides a built-in benchmark mode that disables output writing and enables asynchronous timing, similar to our extension of the Poisson application.

\section{Results}\label{sec:results}

The results presented in this work can be separated into two parts.
Firstly, we use the lightweight Poisson application to expose the overheads of the Legion and HPX backends compared to the MPI backend.
On a single node, we conduct a problem size benchmark to showcase the intra-node task management overhead of the runtimes.
Furthermore, we conduct a strong and weak scaling benchmark to focus on inter-node communication and investigate the scalability of both AMTR backends. 
Secondly, we perform strong and weak scaling benchmarks using the HARD application to examine if complex real-world applications can profit from the asynchronous parallelism introduced by the HPX backend.

The hardware specifications of the Chicoma supercomputer are provided in Table \ref{tab:specs}.
All backends use their respective fastest combination of processes and threads per process, denoted per node as $(\#processes/\#threads)$. 
Naturally, this results in the combination $(128/1)$ for pure MPI. 
The combinations for the other backends were manually tuned for Chicoma. 
The chiplet design of the AMD Zen 2 CPU divides the die into four separate quadrants with two Core Complex Dies each. As a result, the optimal configuration for a dual socket node is $(8/16)$.
The Legion backend is a special case, as one thread per process is reserved for managing the runtime. 
Through Kokkos, FleCSI can utilize OpenMP in the MPI and Legion backends for shared-memory parallelization. 
The HPX backend of FleCSI relies on an HPX parallel loop for shared-memory parallelization.
We use thread binding to physical cores to ensure optimal thread distribution for OpenMP and HPX.

For better comparability of the MPI and HPX backends, we use the MPI parcelport of HPX such that both backends use the same MPI implementation for communication. 
Although this reduces the performance of the HPX backend, it ensures that potential performance improvements over the MPI backend are solely related to the asynchronous parallelization introduced by the HPX runtime.
For the Legion backend, we use GASNet for communication, which is stated to be the most mature in the respective Legion Spack package.

\begin{table}[ht]
\centering
        \begin{tabular}{ll}
        \toprule
        Name         & Chicoma                   \\ \midrule
        \rowcolor{lightgray!50}
        Rpeak           & 9.54PFLOPS                \\ %\hline%
        %\rowcolor{lightgray}
        Nodes           & 1792                      \\ %\hline
        \rowcolor{lightgray!50}
        Connection      & Slingshot-11              \\ %\hline
        %\rowcolor{lightgray}
        Speed           & 200Gb/s                   \\ %\hline
        \rowcolor{lightgray!50}
        Sockets         & 2                         \\ %\hline
        %\rowcolor{lightgray}
        CPU             & AMD EPYC 7H12             \\ %\hline
        \rowcolor{lightgray!50}
        Cores           & 64                        \\ %\hline
        %\rowcolor{lightgray}
        Base clock      & 2.6GHz                    \\ %\hline
        \rowcolor{lightgray!50}
        L3 Cache        & 256MB                     \\ %\hline
        %\rowcolor{lightgray}
        RAM             & 512GB                     \\ 
        \bottomrule
        \end{tabular}
        \vspace{0.1cm}
    \caption{Hardware specification of the Chicoma supercomputer}
    \label{tab:specs}
\end{table}

\subsubsection{Poisson solver}

\begin{figure}
        \centering
        %\raggedleft
        \includegraphics[width=\linewidth]{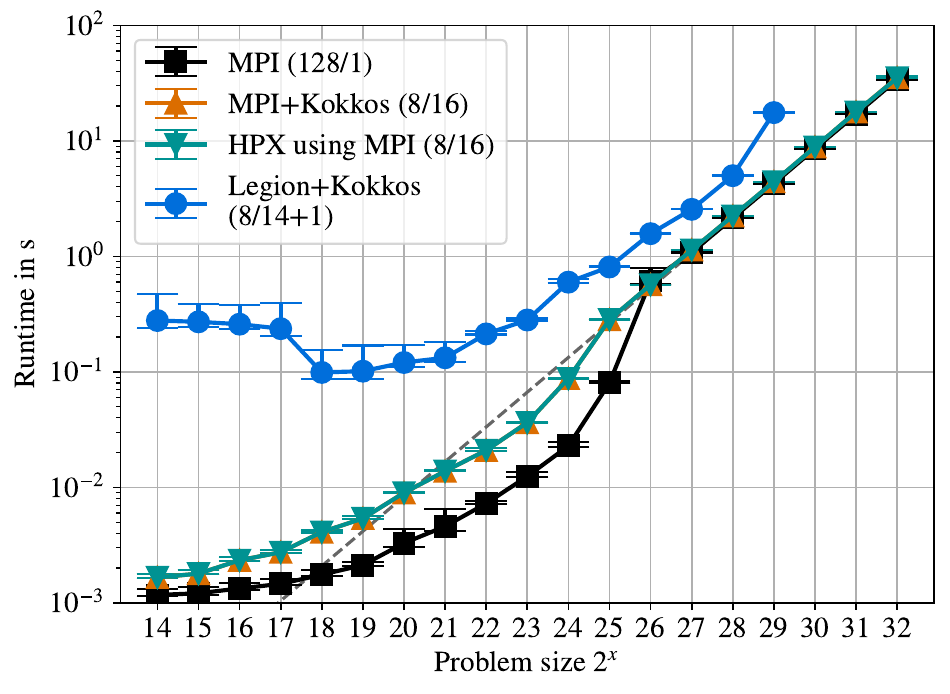}
        \caption{Problem size scaling of one Poisson solver iteration on one node. 
        For Legion, the runtimes do not include tracing optimization overhead.
        Linear scaling is indicated with a dashed line.
        The run configuration is denoted with $(\#process/\#threads)$.} 
        \label{fig:size_poisson}
\end{figure}

Starting with the Poisson application, we conduct scaling tests on the Chicoma supercomputer on up to 1024 nodes.
All results for the Poisson solver are based on averaged runtimes computed from ten runs and five iterations per run.
Error bars indicate the 95\% confidence interval.
To put the scaling tests into perspective, we first present the result of a problem size benchmark on one Chicoma node (see Figure \ref{fig:size_poisson}).
For Legion, the first three iterations, related to trace optimization, are not considered.
For a problem size greater than $2^{25}$, MPI, MPI+Kokkos, and HPX perform similarly. 
However, for problem sizes of $2^{25}$ and smaller, MPI has a clear performance advantage. 
The MPI+Kokkos and HPX backends scale similarly across all problem sizes, indicating the overhead compared to pure MPI is caused by the shared-memory fork-join approach of the computation kernel.
For smaller problem sizes, the larger overheads of the Legion backend become visible. 
Furthermore, Legion requires more memory, thereby reducing the maximum problem size on a single node. 
Note that the scaling is non-linear, which can be explained by more efficient cache usage for smaller problem sizes. 
This introduces non-linearity into the following strong scaling benchmark.

\begin{figure}
        \centering
        %\raggedright
        \includegraphics[width=\linewidth]{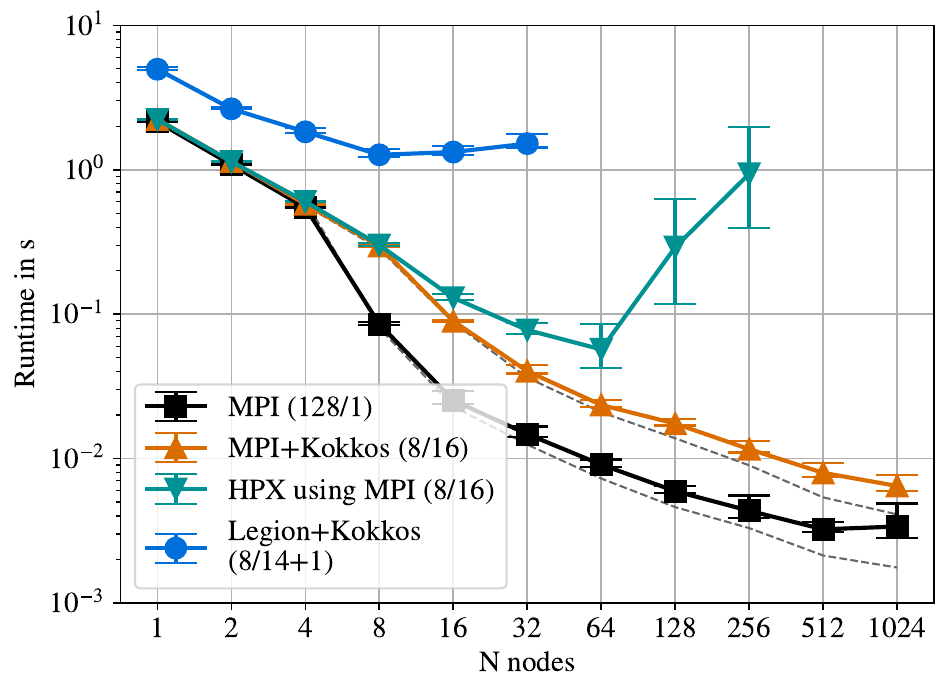}
        \caption{Strong scaling runtimes of one Poisson solver iteration on up to 1024 nodes. 
        The problem size is fixed to $2^{28}$. 
        Ideal single-node performance is indicated with dashed lines.
        The run configuration per node is denoted with $(\#process/\#threads)$.}
        \label{fig:strong_poisson}
\end{figure}

The full-scale strong scaling benchmark is conducted with a problem size of $2^{28}$. 
The resulting runtimes on up to 1024 nodes are presented in Figure \ref{fig:strong_poisson}. 
Optimal single-node performance is indicated with dashed lines.
As the size scaling benchmark in Figure \ref{fig:size_poisson} suggests, the pure MPI backend delivers the best performance. 
The MPI+Kokkos backend performs as expected, with non-linear scaling starting at half the problem size per node, compared to the MPI backend. 
For 1024 nodes, the MPI and MPI+Kokkos backends achieve parallel efficiencies of 55\% and 64\%, respectively.
The better efficiency of the MPI+Kokkos backend is related to the fact that it requires 16 times fewer MPI ranks that participate in communication.

In contrast, the HPX backend maintains ideal strong scaling only up to eight nodes. 
Beyond this point, its scaling declines up to 64 nodes, and performance deteriorates significantly for larger configurations.
We dial this behavior down to the fact that FleCSI relies on collective operations. 
In the HPX backend, those collectives are mapped to the respective HPX collective implementations.
In contrast to MPI collectives, these are currently not implemented with efficient tree-based algorithms. 
In current HPX releases, collectives are implemented such that one host process serves as a communication handler. 
This bottleneck results in significant overhead for large numbers of processes. 
The HPX development team is currently working on making the HPX implementations of the collective operations more efficient, and plans to present preliminary results in a future publication. 

\begin{figure}
        \centering
        %\raggedleft
        \includegraphics[width=\linewidth]{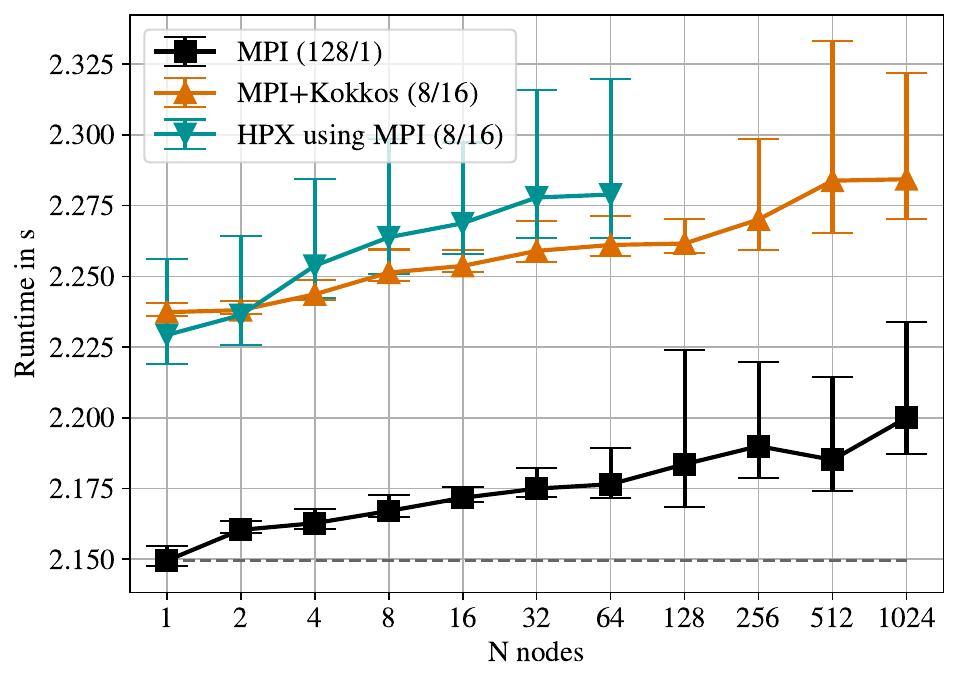}
        \caption{Weak scaling runtimes of one Poisson solver iteration on up to 1024 nodes. 
        The problem size per node is set to $2^{28}$. 
        Ideal scaling is indicated with a dashed line.
        The run configuration per node is denoted with $(\#process/\#threads)$.}
        \label{fig:weak_poisson}
\end{figure}

Regarding weak scaling, we perform a large-scale benchmark on up to 1024 nodes for a problem size of $2^{28}$ per node (see Figure \ref{fig:weak_poisson}). 
The MPI-based backends demonstrate nearly optimal efficiency, achieving over 97\% parallel efficiency on 1024 nodes.
Pure MPI shows a slight performance advantage over the threading backends of roughly 4\%.
The HPX backend exhibits higher variances and slightly worse scaling than the MPI-based backends.
For more than 64 nodes, HPX suffers from the same limitations as in the strong scaling benchmark, such that we exclude performance numbers for higher node counts.
Due to the time-consuming Legion tracing in a distributed setting, we are not able to present Legion performance numbers in this test.

In our Poisson benchmarks, both AMTR backends by design show measurable overheads relative to pure MPI.
However, our analysis also reveals unexpected scaling limitations.
For the HPX backend, a primary bottleneck arises from the collective communication algorithms in HPX.
Apart from this internal issue, HPX delivers performance comparable to the MPI+Kokkos backend, indicating small runtime overheads.
In contrast, for the Legion backend, we were unable to precisely identify the sources of the large runtime overhead, memory footprint, and reduced scaling efficiency.

\subsubsection{HARD}

\begin{figure}
        \centering
        %\raggedright
        \includegraphics[width=\linewidth]{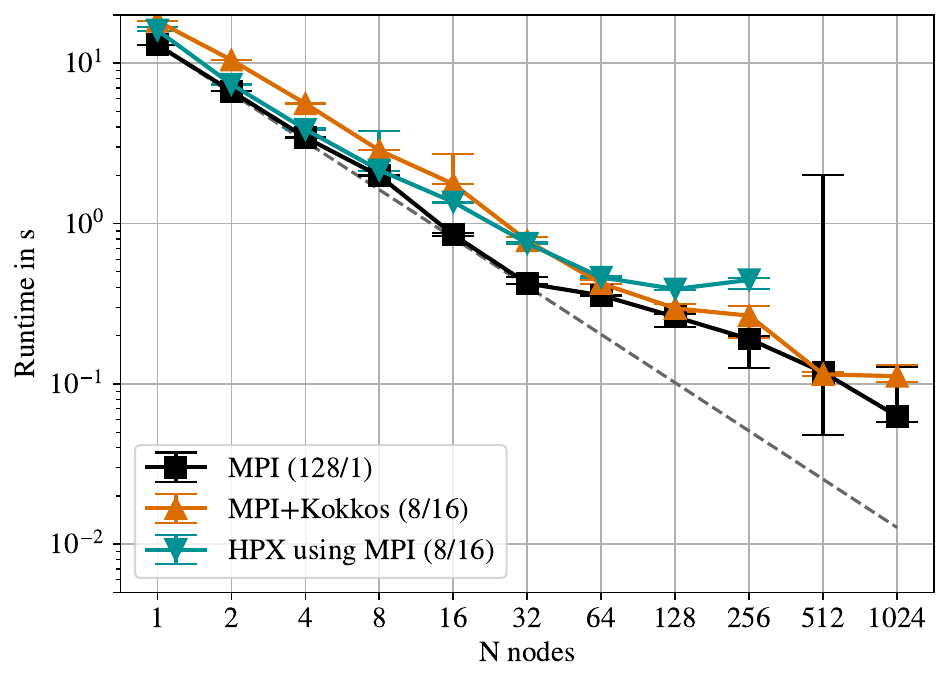}
        \caption{Strong scaling runtimes of one iteration of the three-dimensional radiation hydrodynamics benchmark with HARD on up to 1024 Chicoma nodes (see Figure \ref{tab:specs}). The problem size is fixed to $2^{9+9+9}$. 
        Ideal scaling is indicated with a dashed line. The run configuration per node is denoted with $(\#process/\#threads)$.}
        \label{fig:strong_hard_rad}
\end{figure}

For the HARD application, all presented results are based on the median iteration runtime from five runs.
Error bars indicate the minimum and maximum of these five runs.
We directly start with a strong scaling benchmark up to 1024 nodes for a three-dimensional problem size of $2^{9+9+9}$ (see Figure \ref{fig:strong_hard_rad}).
Due to a more compute-heavy and less communication-focused workload, the scaling limitations of the HPX backend can be mitigated.
While the pure MPI backend again shows the best performance and optimal scaling up to 64 cores, the HPX backend now exhibits superior performance compared to the MPI+Kokkos backend on up to 32 nodes. 
For up to 16 nodes, we observe speedups ranging from $1.14$ to $1.45$ for HPX over MPI+Kokkos.

\begin{figure}
        \centering
        %\raggedright
        \includegraphics[width=\linewidth]{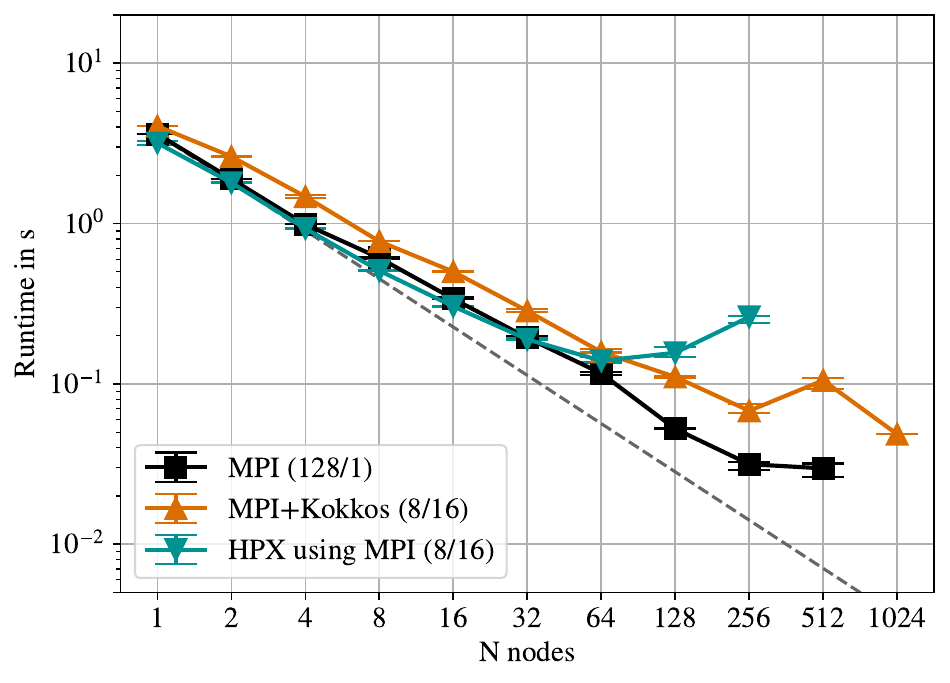}
        \caption{Strong scaling runtimes of one iteration of the three-dimensional hydrodynamics-only benchmark with HARD on up to 1024 Chicoma nodes (see Figure \ref{tab:specs}). The problem size is fixed to $2^{9+9+9}$. 
        Ideal scaling is indicated with a dashed line. The run configuration per node is denoted with $(\#process/\#threads)$.}
        \label{fig:strong_hard_hydro}
\end{figure}

If we disable radiation and only compute the hydrodynamics part of our strong scaling benchmark with HARD, we can observe a single-node performance advantage of the HPX backend in Figure \ref{fig:strong_hard_hydro}.
The HPX backend shows the best single-node performance, showing a speedup of $1.14$ over the MPI backend and a speedup of $1.27$ over MPI+Kokkos.
Additionally, the HPX backend maintains its performance advantage up to 64 nodes until its scaling limitations take over.
For 32 or fewer nodes, HPX shows a speedup over MPI+Kokkos of up to $1.64$ and over MPI of up to $1.20$.

\begin{figure}
        \centering
        %\raggedright
        \includegraphics[width=\linewidth]{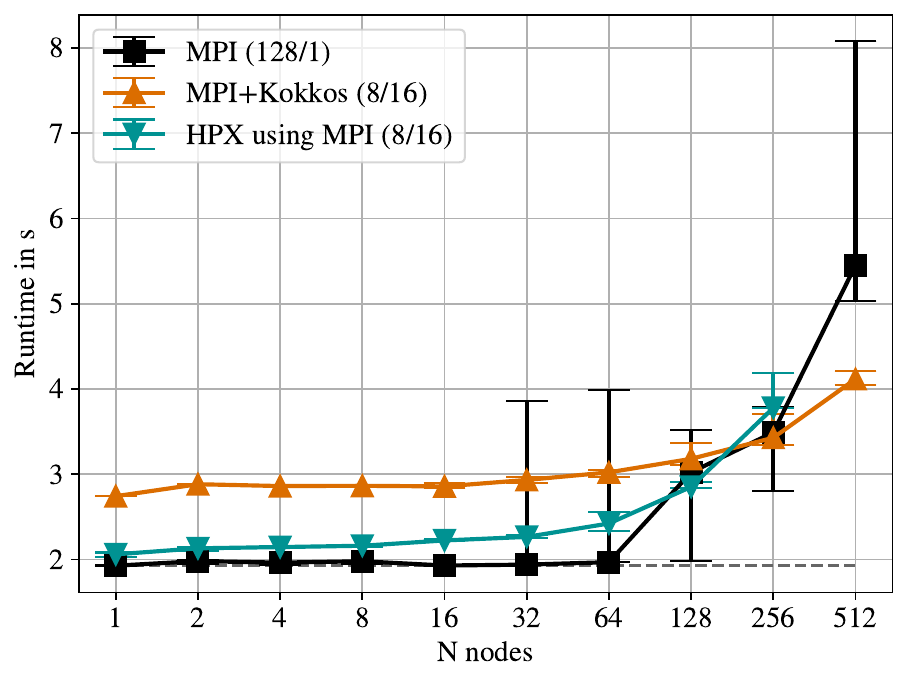}
        \caption{Weak scaling runtimes of one iteration of the three-dimensional radiation hydrodynamics benchmark with HARD on up to 512 Chicoma nodes (see Figure \ref{tab:specs}). The problem size per node is fixed to $2^{8+8+8}$. 
        Ideal scaling is indicated with a dashed line. The run configuration per node is denoted with $(\#process/\#threads)$.}
        \label{fig:weak_hard_rad}
\end{figure}

We present the results of the HARD weak scaling benchmark up to 512 nodes for the three-dimensional radiation problem in Figure~\ref{fig:weak_hard_rad}. 
The problem size per node is fixed to $2^{8+8+8}$.
Although pure MPI delivers ideal efficiency up to 64 nodes, performance starts to rapidly decrease.
In contrast, MPI+Kokkos, which uses 16 times fewer MPI ranks, achieves better scaling.
Between a single and 64 nodes, HPX shows an average speedup of $1.31$ over the MPI+Kokkos approach.
Both backends rely on asynchronous MPI for communication and use a fork-join approach for large computation kernels. 
However, the HPX backend can utilize the HPX runtime's ability to asynchronously schedule computation kernels. 

\section{Conclusion and outlook}\label{sec:conclusion}

In this work, we evaluated the performance and scalability of the FleCSI framework and its backends on a large scale using
an iterative Poisson solver and the radiation hydrodynamics code HARD on up to 1024 nodes on the Chicoma supercomputer.
We developed a benchmark mode for the FleCSI Poisson example that enables asynchronous timing and facilitates accurate performance comparisons across different backends.
Leveraging FleCSI’s backend architecture, we analyzed the runtime overheads of two popular AMTRs relative to traditional MPI execution.

For the Poisson application, FleCSI achieved a parallel efficiency of more than 97\% on up to 131072 cores in our weak scaling benchmark.
Combining MPI with Kokkos resulted in slightly lower performance due to fork–join kernel overheads; however, FleCSI demonstrated improved scalability as a result of reduced communication costs.
Within the more complex radiation-hydrodynamics benchmark using HARD, the HPX backend showcased the advantages of its asynchronous execution model.
Up to 64 nodes, the HPX backend outperformed the MPI+Kokkos backend up to a factor of $1.27$ in strong scaling and on an average factor of $1.31$ in weak scaling. 
For our hydrodynamics-only strong scaling benchmark, the HPX backend achieved speedups of up to $1.64$ over MPI+Kokkos and up to $1.20$ over pure MPI.

Addressing our formulated research questions, the Poisson benchmark results demonstrate that FleCSI introduces only a lightweight abstraction layer on top of conventional MPI-based parallelization. 
Both AMTRs incur measurable runtime overhead and exhibit scaling limitations compared to the MPI-based implementation.
However, the HPX runtime shows lower overhead than the Legion runtime.
In contrast, the HARD benchmark results highlight the advantages of AMTRs for complex scientific simulations, where asynchronous interleaving of computation and communication can improve resource utilization and overall performance.

Several open questions remain for future investigation.
In particular, the performance of the Legion backend requires further analysis.
We plan to profile the Legion backend using the Poisson application to identify and quantify the origins of the runtime overheads and increased memory usage.
Regarding the scaling limitations of the HPX backend, our analysis indicated that a primary bottleneck lies in the usage of HPX collectives. 
Consequently, we intend to update our scaling benchmarks once optimized HPX collectives become available.
Finally, we plan to extend our benchmarks to heterogeneous architectures that include GPU accelerators.
In such environments, the AMTR backends may offer even greater performance advantages over the synchronous MPI backend by overlapping computation and CPU-GPU data transfers.

\section*{Acknowledgments}
The benchmarks in this work were conducted as part of the Co-Design Summer School program hosted by Los Alamos National Laboratory (LANL).
The authors want to specifically thank the summer school hosts J. Loiseau and H. Lim, as well as E. Maestas, C. Junghans, K. Thompson, and our
mentors, B. Bergen, R. Berger, P. Edelmann, B. Krueger, S. Lakshmiranganatha, A. Y. López, M.
Moraru, N. Prajapati, A. Reisner, and T. Vogel.
Computations were performed on the Chicoma supercomputer through the Advanced Simulation and Computing (ASC) program at LANL. 
The data is approved for public release under LA-UR-24-28836.

\section*{Supplementary materials}
The FleCSI code is publicly available on \href{https://github.com/flecsi/flecsi}{Github}\footnote{\url{https://github.com/flecsi/flecsi} accessed: 2026-01-10}. FleCSI can be compiled with MPI, Legion, and Kokkos support for version 2.3. The exact implementation of the HPX backend used in this work is not part of an official release. However, a refined version of the HPX backend is available in FleCSI version 2.4.
For our radiation hydrodynamics benchmarks, we used HARD version 1.0, which is publicly available on \href{https://github.com/lanl/HARD}{Github}\footnote{\url{https://github.com/lanl/HARD} accessed: 2026-01-10}.

Our benchmarks were conducted with a modified version of the Poisson example. This benchmark version disables I/O and allows one to time the performance of the Legion and HPX backends asynchronously.  
The corresponding git patch for the FleCSI example, Spack environments containing the versions of all used software, scenario configuration files, and benchmark scripts for strong and weak scaling are available at \href{https://doi.org/10.18419/DARUS-5193}{DaRUS}\footnote{\url{https://doi.org/10.18419/DARUS-5193} accessed: 2026-03-03}.

\section*{AI Usage Disclosure}

Generative artificial intelligence (AI) tools, including Grammarly~\cite{grammarly}, Writefull~\cite{writefull}, and ChatGPT~\cite{chatgpt}, were employed to enhance the clarity, grammar, and overall coherence of the manuscript. All technical content, data analyses, and research findings were conceived and developed independently by the authors. AI-assisted outputs were carefully reviewed, verified, and edited by the authors to ensure factual accuracy, interpretive rigor, and scholarly integrity. The final manuscript reflects the authors’ original intellectual contributions and analytical work.

\bibliographystyle{IEEEtran}
\bibliography{main}

@misc{Bergen2016_flecsi_1,
  author       = {Bergen, Ben and Moss, Nicholas and Charest, Marc Robert Joseph},
  title        = {Flexible Computer Science Infrastructure (FleCSI)},
  url          = {https://www.osti.gov/biblio/1311634},
  place        = {United States},
  year         = {2016},
  month        = {04}}

@InProceedings{Bergen2022_flecsi,
author="Bergen, Ben
and Demeshko, Irina
and Ferenbaugh, Charles
and Herring, Davis
and Lo, Li-Ta
and Loiseau, Julien
and Ray, Navamita
and Reisner, Andrew",
title="FleCSI 2.0: The Flexible Computational Science Infrastructure Project",
booktitle="Euro-Par 2021: Parallel Processing Workshops",
year="2022",
publisher="Springer International Publishing",
address="Cham",
pages="480--495",
isbn="978-3-031-06156-1"
}

@inproceedings{Holmen2021_uintah,
  author    = {J.K. Holmen and D. Sahasrabudhe and M. Berzins},
  title     = {A heterogeneous MPI+PPL task scheduling approach for asynchronous many-task runtime systems},
  booktitle = {Proceedings of the Practice and Experience in Advanced Research Computing 2021 on Sustainability, Success and Impact (PEARC21)},
  year      = {2021},
  publisher = {ACM},
}

@inproceedings{Perache2009_mpc-mpi,
  author    = {M. P\'erache and P. Carribault and H. Jourdren},
  title     = {{MPC-MPI}: an {MPI} implementation reducing the overall memory consumption},
  booktitle = {EuroPVM/MPI 2009},
  editor    = {M. Ropo and J. Westerholm and J. Dongarra},
  series    = {Lecture Notes in Computer Science},
  volume    = {5759},
  pages     = {94--103},
  publisher = {Springer},
  address   = {Heidelberg},
  year      = {2009},
  doi       = {10.1007/978-3-642-03770-2_16}
}

@article{Loiseau2020_flecsph,
title = {FleCSPH: The next generation FleCSIble parallel computational infrastructure for smoothed particle hydrodynamics},
journal = {SoftwareX},
volume = {12},
pages = {100602},
year = {2020},
issn = {2352-7110},
doi = {https://doi.org/10.1016/j.softx.2020.100602},
url = {https://www.sciencedirect.com/science/article/pii/S2352711020303150},
author = {Julien Loiseau and Hyun Lim and Mark Alexander Kaltenborn and Oleg Korobkin and Christopher M. Mauney and Irina Sagert and Wesley P. Even and Benjamin K. Bergen},
}

@inproceedings{Boehme2016_caliper,
  author={Boehme, David and Gamblin, Todd and Beckingsale, David and Bremer, Peer-Timo and Gimenez, Alfredo and LeGendre, Matthew and Pearce, Olga and Schulz, Martin},

  title     = {Caliper: Performance Introspection for HPC Software Stacks},
  booktitle = {Proceedings of the International Conference for High Performance Computing, Networking, Storage and Analysis (SC16)},
  pages     = {550--560},
  year      = {2016},
  doi       = {10.1109/SC.2016.46}
}

@Article{Thoman2018_taxonomy,
  author    = {Thoman, Peter and Dichev, Kiril and Heller, Thomas and Iakymchuk, Roman and Aguilar, Xavier and Hasanov, Khalid and Gschwandtner, Philipp and Lemarinier, Pierre and Markidis, Stefano and Jordan, Herbert and Fahringer, Thomas and Katrinis, Kostas and Laure, Erwin and Nikolopoulos, Dimitrios S.},
  journal   = {J. Supercomput.},
  title     = {A Taxonomy of Task-Based Parallel Programming Technologies for High-Performance Computing},
  year      = {2018},
  issn      = {0920-8542},
  month     = {04},
  number    = {4},
  pages     = {1422–1434},
  volume    = {74},
  address   = {USA},
  doi       = {10.1007/s11227-018-2238-4},
  numpages  = {13},
  publisher = {Kluwer Academic Publishers},
}

@INPROCEEDINGS{Slaughter2020_taskbench,
  author={Slaughter, Elliott and Wu, Wei and Fu, Yuankun and Brandenburg, Legend and Garcia, Nicolai and Kautz, Wilhem and Marx, Emily and Morris, Kaleb S. and Cao, Qinglei and Bosilca, George and Mirchandaney, Seema and Leek, Wonchan and Treichlerk, Sean and McCormick, Patrick and Aiken, Alex},
  booktitle={SC20: International Conference for High Performance Computing, Networking, Storage and Analysis}, 
  title={Task Bench: A Parameterized Benchmark for Evaluating Parallel Runtime Performance}, 
  year={2020},
  volume={},
  number={},
  pages={1-15},
  doi={10.1109/SC41405.2020.00066}}

@InCollection{Wu2023_taskbench_hpx,
  author    = {Nanmiao Wu and Ioannis Gonidelis and Simeng Liu and Zane Fink and Nikunj Gupta and Karame Mohammadiporshokooh and Patrick Diehl and Hartmut Kaiser and Laxmikant V. Kale},
  booktitle = {Euro-Par 2022: Parallel Processing Workshops},
  publisher = {Springer Nature Switzerland},
  title     = {Quantifying Overheads in Charm++ and {HPX} Using Task Bench},
  year      = {2023},
  pages     = {5-16},
  doi       = {10.1007/978-3-031-31209-0_1},
}

@INPROCEEDINGS{Bauer2012_legion,
  author={Bauer, Michael and Treichler, Sean and Slaughter, Elliott and Aiken, Alex},
  booktitle={SC '12: Proceedings of the International Conference on High Performance Computing, Networking, Storage and Analysis}, 
  title={Legion: Expressing locality and independence with logical regions}, 
  year={2012},
  volume={},
  number={},
  pages={1-11},
  doi={10.1109/SC.2012.71}}

@INPROCEEDINGS{Treichler2014_realm,
  author={Treichler, Sean and Bauer, Michael and Aiken, Alex},
  booktitle={2014 23rd International Conference on Parallel Architecture and Compilation Techniques (PACT)}, 
  title={Realm: An event-based low-level runtime for distributed memory architectures}, 
  year={2014},
  volume={},
  number={},
  pages={263-275},
  doi={10.1145/2628071.2628084}}

@inproceedings{Slaugther2015_regent,
author = {Slaughter, Elliott and Lee, Wonchan and Treichler, Sean and Bauer, Michael and Aiken, Alex},
title = {Regent: a high-productivity programming language for HPC with logical regions},
year = {2015},
isbn = {9781450337236},
publisher = {Association for Computing Machinery},
address = {New York, NY, USA},
url = {https://doi.org/10.1145/2807591.2807629},
doi = {10.1145/2807591.2807629},
booktitle = {Proceedings of the International Conference for High Performance Computing, Networking, Storage and Analysis},
articleno = {81},
numpages = {12},
location = {Austin, Texas},
series = {SC '15}
}

@Article{Marcello2021_octotiger,
  author    = {Dominic C Marcello and Sagiv Shiber and Orsola De~Marco and Juhan Frank and Geoffrey C Clayton and Patrick M Motl and Patrick Diehl and Hartmut Kaiser},
  journal   = {Monthly Notices of the Royal Astronomical Society},
  title     = {{Octo-Tiger: a new, 3D hydrodynamic code for stellar mergers that uses HPX parallelization}},
  year      = {2021},
  month     = {04},
  number    = {4},
  pages     = {5345-5382},
  volume    = {504},
  comment   = {Summary: tbd.},
  doi       = {10.1093/mnras/stab937},
  publisher = {Oxford University Press ({OUP})},
}

@InProceedings{Daiss2022_hpx_kokkos,
  author    = {Gregor Daiß and Srinivas Yadav Singanaboina and Patrick Diehl and Hartmut Kaiser and Dirk Pfluger},
  booktitle = {2022 {IEEE}/{ACM} 7th International Workshop on Extreme Scale Programming Models and Middleware ({ESPM}2)},
  title     = {From Merging Frameworks to Merging Stars: Experiences using {HPX}, Kokkos and {SIMD} Types},
  year      = {2022},
  month     = {11},
  publisher = {{IEEE}},
  doi       = {10.1109/espm256814.2022.00007},
}

@InProceedings{Daiss2023_hpx_sycl,
  author    = {Dai\ss{}, Gregor and Diehl, Patrick and Kaiser, Hartmut and Pfl\"{u}ger, Dirk},
  booktitle = {Proceedings of the 2023 International Workshop on OpenCL},
  title     = {{Stellar Mergers with HPX-Kokkos and SYCL: Methods of Using an Asynchronous Many-Task Runtime System with SYCL}},
  year      = {2023},
  address   = {New York, NY, USA},
  publisher = {Association for Computing Machinery},
  series    = {IWOCL '23},
  articleno = {8},
  doi       = {10.1145/3585341.3585354},
  isbn      = {9798400707452},
  location  = {Cambridge, United Kingdom},
  numpages  = {12},
}

@InProceedings{Helmann2025_gprat,
author="Helmann, Maksim
and Strack, Alexander
and Pfl{\"u}ger, Dirk",
editor="Diehl, Patrick
and Cao, Qinglei
and Herault, Thomas
and Bosilca, George",
title="GPRat: Gaussian Process Regression with Asynchronous Tasks",
booktitle="Asynchronous Many-Task Systems and Applications",
year="2026",
publisher="Springer Nature Switzerland",
address="Cham",
pages="83--94",
isbn="978-3-031-97196-9"
}

@article{Kaiser2020_hpx,
    doi = {10.21105/joss.02352}, 
    url = {https://doi.org/10.21105/joss.02352},
    year = {2020}, 
    publisher = {The Open Journal}, 
    volume = {5}, 
    number = {53}, 
    pages = {2352}, 
    author = {Kaiser, Hartmut and Diehl, Patrick and Lemoine, Adrian S. and Lelbach, Bryce Adelstein and Amini, Parsa and Berge, Agustín and Biddiscombe, John and Brandt, Steven R. and Gupta, Nikunj and Heller, Thomas and Huck, Kevin and Khatami, Zahra and Kheirkhahan, Alireza and Reverdell, Auriane and Shirzad, Shahrzad and Simberg, Mikael and Wagle, Bibek and Wei, Weile and Zhang, Tianyi}, 
    title = {{HPX - The C++ Standard Library for Parallelism and Concurrency}}, 
    journal = {Journal of Open Source Software} }

@book{Groop1999_mpich,
  author    = {William Gropp and Ewing Lusk and Anthony Skjellum},
  title     = {Using MPI: Portable Parallel Programming with the Message Passing Interface},
  publisher = {MIT Press},
  edition   = {2},
  year      = {1999}
}

@InProceedings{Gabriel2004_openmpi,
  author="Gabriel, Edgar
and Fagg, Graham E.
and Bosilca, George
and Angskun, Thara
and Dongarra, Jack J.
and Squyres, Jeffrey M.
and Sahay, Vishal
and Kambadur, Prabhanjan
and Barrett, Brian
and Lumsdaine, Andrew
and Castain, Ralph H.
and Daniel, David J.
and Graham, Richard L.
and Woodall, Timothy S.",
editor="Kranzlm{\"u}ller, Dieter
and Kacsuk, P{\'e}ter
and Dongarra, Jack",
  booktitle = {Proceedings, 11th European PVM/MPI Users' Group Meeting},
  title     = {Open {MPI}: Goals, Concept, and Design of a Next Generation {MPI} Implementation},
  year      = {2004},
  address   = {Budapest, Hungary},
  month     = {September},
  pages     = {97--104},
}

@manual{Mpi2021,
    author = "{Message Passing Interface Forum}",
    title  = "{{MPI}: A Message-Passing Interface Standard Version 3.0}",
    year   = 2021
}

@InProceedings{Yan23_lci,
  author    = {Yan, Jiakun and Kaiser, Hartmut and Snir, Marc},
  booktitle = {Proceedings of the SC '23 Workshops},
  title     = {{Design and Analysis of the Network Software Stack of an Asynchronous Many-task System -- The LCI parcelport of HPX}},
  year      = {2023},
  address   = {New York, NY, USA},
  pages     = {1151–1161},
  publisher = {ACM},
  isbn      = {9798400707858},
  location  = {, Denver, CO, USA,},
  numpages  = {11},
}

@PhdThesis{Heller2019_agas,
  author   = {Heller, Thomas},
  school   = {FAU Erlangen-Nürnberg},
  title    = {{Extending the C++ Asynchronous Programming Model with the HPX Runtime System for Distributed Memory Computing}},
  year     = {2019},
}

@article{Trott2022_kokkos,
  author={Trott, Christian R. and Lebrun-Grandié, Damien and Arndt, Daniel and Ciesko, Jan and Dang, Vinh and Ellingwood, Nathan and Gayatri, Rahulkumar and Harvey, Evan and Hollman, Daisy S. and Ibanez, Dan and Liber, Nevin and Madsen, Jonathan and Miles, Jeff and Poliakoff, David and Powell, Amy and Rajamanickam, Sivasankaran and Simberg, Mikael and Sunderland, Dan and Turcksin, Bruno and Wilke, Jeremiah},
  journal={IEEE Transactions on Parallel and Distributed Systems},
  title={Kokkos 3: Programming Model Extensions for the Exascale Era},
  year={2022},
  volume={33},
  number={4},
  pages={805-817},
  doi={10.1109/TPDS.2021.3097283}}

@misc{cray-mpich,
  author       = {{Cray Inc.}},
  title        = {{Cray MPICH}},
  year         = {2025},
  howpublished = {\url{https://cpe.ext.hpe.com/docs/24.03/mpt/mpich/index.html#mpich}},
  note         = {Version 8.1.26. Accessed: 2025-07-14}
}

@book{Saad2003_sparse_methods,
author = {Saad, Y.},
title = {Iterative Methods for Sparse Linear Systems},
year = {2003},
isbn = {0898715342},
publisher = {Society for Industrial and Applied Mathematics},
address = {USA},
edition = {2nd}
}

@techreport{Bonachea2017_gasnet,
  author    = {Bonachea, D. and Hargrove, P. H.},
  title     = {{GASNet specification, v1.8.1}},
  institution = {Lawrence Berkeley National Laboratory},
  year      = {2017},
  month     = aug,
  number    = {LBNL-2001064},
  doi       = {10.2172/1398512},
  note      = {\url{https://doi.org/10.2172/1398512}}
}

@inproceedings{Diehl2023_riscv,
    author = {Diehl, Patrick and Daiss, Gregor and Brandt, Steven and Kheirkhahan, Alireza and Kaiser, Hartmut and Taylor, Christopher and Leidel, John},
    title = {{Evaluating HPX and Kokkos on RISC-V using an astrophysics application Octo-Tiger}},
    year = {2023},
    isbn = {9798400707858},
    publisher = {Association for Computing Machinery},
    address = {New York, NY, USA},
    booktitle = {SC-W 2023},
    pages = {1533–1542},
    numpages = {10},
    location = {Denver, CO, USA},
}

@misc{Diehl2023_arm,
author = {Diehl, Patrick and Daiß, Gregor and Huck, Kevin and Marcello, Dominic and Shiber, Sagiv and Kaiser, Hartmut and Pflüger, Dirk},
year = {2023},
month = {03},
pages = {},
title = {{Simulating Stellar Merger using HPX/Kokkos on A64FX on Supercomputer Fugaku}},
doi = {10.48550/arXiv.2304.11002}
}

@article{Turner2001_fld_numerical,
  archiveprefix = {arXiv},
  author        = {Turner, N. J. and Stone, J. M.},
  doi           = {10.1086/321779},
  eprint        = {astro-ph/0102145},
  journal       = {Astrophys. J. Suppl.},
  pages         = {95--108},
  title         = {{A module for radiation hydrodynamic calculations with zeus-2d using flux-limited diffusion}},
  volume        = {135},
  year          = {2001}
}

@ARTICLE{Levermore1981_fld_original,
       author = {{Levermore}, C.~D. and {Pomraning}, G.~C.},
        title = "{A flux-limited diffusion theory}",
      journal = {The Astrophysical Journal},
         year = 1981,
        month = aug,
       volume = {248},
        pages = {321-334},
          doi = {10.1086/159157}
}

@inbook{Strack2024_hpxfft,
   title={{Experiences Porting Shared and Distributed Applications to Asynchronous Tasks: A Multidimensional FFT Case-Study}},
   ISBN={9783031617638},
   ISSN={1611-3349},
   booktitle={Asynchronous Many-Task Systems and Applications},
   publisher={Springer Nature Switzerland},
   author={Strack, Alexander and Taylor, Christopher and Diehl, Patrick and Pflüger, Dirk},
   year={2024},
   pages={111–122}
}

@misc{Cerf1974_tcp,
    author = {Cerf, Vinton G. and Dalal, Yogen K.},
    series =    {Request for Comments},
    number =    675,
    howpublished =  {RFC 675},
    publisher = {RFC Editor},
    title =     {{Specification of Internet Transmission Control Program}},
    pagetotal = 70,
    year =      1974,
}

@inproceedings{Strack2025_parcelports,
author = {Strack, Alexander and Pfl\"{u}ger, Dirk},
title = {{A HPX Communication Benchmark: Distributed FFT Using Collectives}},
year = {2025},
isbn = {978-3-031-90202-4},
publisher = {Springer-Verlag},
address = {Berlin, Heidelberg},
url = {https://doi.org/10.1007/978-3-031-90203-1_25},
doi = {10.1007/978-3-031-90203-1_25},
booktitle = {Euro-Par 2024: Parallel Processing Workshops: Euro-Par 2024 International Workshops, Madrid, Spain, August 26–30, 2024, Proceedings, Part II},
pages = {271–274},
numpages = {4},
location = {Madrid, Spain}
}

@InProceedings{Mirchandaney2024_spinifel_legion,
author="Mirchandaney, Seema
and Aiken, Alex
and Slaughter, Elliott",
editor="Diehl, Patrick
and Schuchart, Joseph
and Valero-Lara, Pedro
and Bosilca, George",
title="Speaking Pygion: Experiences Writing an Exascale Single Particle Imaging Code",
booktitle="Asynchronous Many-Task Systems and Applications",
year="2024",
publisher="Springer Nature Switzerland",
address="Cham",
pages="1--8",
isbn="978-3-031-61763-8"
}

@article{Ferenbaugh2014_pennant,
author = {Ferenbaugh, Charles},
year = {2014},
month = {10},
pages = {},
title = {PENNANT: An unstructured mesh mini-app for advanced architecture research},
volume = {27},
journal = {Concurrency and Computation: Practice and Experience},
doi = {10.1002/cpe.3422}
}

@INPROCEEDINGS{Shamis2015_ucx,
  author={Shamis, Pavel and Venkata, Manjunath Gorentla and Lopez, M. Graham and Baker, Matthew B. and Hernandez, Oscar and Itigin, Yossi and Dubman, Mike and Shainer, Gilad and Graham, Richard L. and Liss, Liran and Shahar, Yiftah and Potluri, Sreeram and Rossetti, Davide and Becker, Donald and Poole, Duncan and Lamb, Christopher and Kumar, Sameer and Stunkel, Craig and Bosilca, George and Bouteiller, Aurelien},
  booktitle={2015 IEEE 23rd Annual Symposium on High-Performance Interconnects}, 
  title={UCX: An Open Source Framework for HPC Network APIs and Beyond}, 
  year={2015},
  volume={},
  number={},
  pages={40-43},
  doi={10.1109/HOTI.2015.13}}

@misc{Chang2021_spinifel,
      title={Scaling and Acceleration of Three-dimensional Structure Determination for Single-Particle Imaging Experiments with SpiniFEL}, 
      author={Hsing-Yin Chang and Elliott Slaughter and Seema Mirchandaney and Jeffrey Donatelli and Chun Hong Yoon},
      year={2021},
      eprint={2109.05339},
      archivePrefix={arXiv},
      primaryClass={physics.comp-ph},
      url={https://arxiv.org/abs/2109.05339}, 
}

@manual{nvidia2025_cuda,
  title        = {CUDA C++ Programming Guide},
  author       = {{NVIDIA Corporation}},
  organization = {NVIDIA},
  year         = {2025},
  note         = {Version 12.9},
  url          = {https://docs.nvidia.com/cuda/cuda-c-programming-guide/}
}

@article{Rankine1870_rk,
 author = {W. J. Macquorn Rankine},
 journal = {Philosophical Transactions of the Royal Society of London},
 pages = {277--288},
 publisher = {Royal Society},
 title = {On the Thermodynamic Theory of Waves of Finite Longitudinal Disturbance},
 volume = {160},
 year = {1870}
}

@article{Hugoniot1887_rk,
  author    = {Hugoniot, H.},
  title     = {{M{\'e}moire sur la propagation du mouvement dans un fluide ind{\'e}fini (premi{\`e}re partie)}},
  journal   = {Journal de math{\'e}matiques pures et appliqu{\'e}es, 4e s{\'e}rie},
  volume    = {3},
  pages     = {477--492},
  year      = {1887}
}

@misc{chatgpt,
  author       = {OpenAI},
  title        = {{ChatGPT 5}},
  year         = {2025},
  howpublished = {\url{https://openai.com/chatgpt}},
  note         = {Accessed: 2026-01-14}
}

@misc{grammarly,
  author       = {{Grammarly, Inc.}},
  title        = {Grammarly},
  year         = {2026},
  howpublished = {\url{https://www.grammarly.com/}},
  note         = {Accessed: 2026-01-14}
}

@misc{writefull,
  author       = {{Digital Science}},
  title        = {Writefull},
  year         = {2025},
  howpublished = {\url{https://www.writefull.com}},
  note         = {Accessed: 2026-01-14}
}

@inproceedings{Herring2025_flecsi_hpx,
author = {Herring, S. Davis and Moraru, Maxim and Pakin, Scott and Loiseau, Julien and Berger, Richard and Edelmann, Philipp V. F. and Bergen, Ben},
title = {{Enhancing HPX with FleCSI: Automatic Detection of Implicit Task Dependencies}},
year = {2025},
isbn = {9798400718717},
publisher = {Association for Computing Machinery},
address = {New York, NY, USA},
url = {https://doi.org/10.1145/3731599.3767507},
doi = {10.1145/3731599.3767507},
booktitle = {Proceedings of the SC '25 Workshops of the International Conference for High Performance Computing, Networking, Storage and Analysis},
pages = {1330–1340},
numpages = {11},
location = {
},
series = {SC Workshops '25}
}

@article{Loiseau2025_hard,
title = {HARD: A performance portable radiation hydrodynamics code based on FleCSI framework},
journal = {SoftwareX},
volume = {32},
pages = {102441},
year = {2025},
issn = {2352-7110},
doi = {https://doi.org/10.1016/j.softx.2025.102441},
url = {https://www.sciencedirect.com/science/article/pii/S2352711025004078},
author = {Julien Loiseau and Hyun Lim and Andrés {Yagüe López} and Mammadbaghir Baghirzade and Shihab Shahriar Khan and Yoonsoo Kim and Sudarshan Neopane and Alexander Strack and Farhana Taiyebah and Ben Bergen},
}

@article{Schuchart2025,
author = {Schuchart, Joseph and Diehl, Patrick and Bauer, Michael and Bouteiller, Aurelien and Daiss, Gregor and Kayraklioglu, Engin and Khandekar, Shreyas and Herault, Thomas and Holmen, John and Rao, Ritvik and Strack, Alexander and Schlaughter, Elliott and Spinti, Jennifer and Thornock, Jeremy and Aiken, Alex and Aumage, Olivier and Berzins, Martin and Bosilca, George and Chamberlain, Bradford L. and Kale, Laxmikant},
year = {2025},
Journal = {Authorea Preprints},
month = {12},
pages = {},
title = {A Survey of Distributed Asynchronous Many-Task Models and Their Applications},
note={Preprint}
}

@InProceedings{Strack2026_riscv,
author="Strack, Alexander
and Taylor, Christopher
and Pfl{\"u}ger, Dirk",
editor="Neuwirth, Sarah
and Paul, Arnab Kumar
and Weinzierl, Tobias
and Carson, Erin Claire",
title="{{Parallel FFTW on RISC-V: A Comparative Study Including OpenMP, MPI, and HPX}}",
booktitle="High Performance Computing",
year="2026",
publisher="Springer Nature Switzerland",
address="Cham",
pages="586--597",
}

@book{Briggs2000_multigrid,
author = {Briggs, William L. and Henson, Van Emden and McCormick, Steve F.},
title = {A Multigrid Tutorial, Second Edition},
publisher = {Society for Industrial and Applied Mathematics},
year = {2000},
doi = {10.1137/1.9780898719505},
address = {},
edition   = {Second},
}

@article{Harten1983_hard_hll_flux,
author = {Harten, Amiram and Lax, Peter D. and Leer, Bram van},
title = {On Upstream Differencing and Godunov-Type Schemes for Hyperbolic Conservation Laws},
journal = {SIAM Review},
volume = {25},
number = {1},
pages = {35-61},
year = {1983},
doi = {10.1137/1025002},
}

@article{Rusanov1962_hard_lax_flux,
title = {The calculation of the interaction of non-stationary shock waves and obstacles},
journal = {USSR Computational Mathematics and Mathematical Physics},
volume = {1},
number = {2},
pages = {304-320},
year = {1962},
issn = {0041-5553},
doi = {https://doi.org/10.1016/0041-5553(62)90062-9},
author = {V.V Rusanov}
}

@article{Borges2008_hard_wenoz,
title = {An improved weighted essentially non-oscillatory scheme for hyperbolic conservation laws},
journal = {Journal of Computational Physics},
volume = {227},
number = {6},
pages = {3191-3211},
year = {2008},
issn = {0021-9991},
doi = {https://doi.org/10.1016/j.jcp.2007.11.038},
author = {Rafael Borges and Monique Carmona and Bruno Costa and Wai Sun Don},
}

\end{document}